\documentclass[11pt,twoside]{article}

\usepackage{asp2006}
\usepackage{lscape}

\begin{document}
\title{Cosmologists in the dark}   
\author{Vicent J. Mart\'{\i}nez$^{1,2}$, and Virginia Trimble$^{3,4}$}
\affil{$^1$ Observatori Astron\`omic de la Universitat de Val\`encia. Ap.
de Correus 22085, E-46071 Val\`encia, Spain\\
$^2$ Departament d'Astronomia i Astrof\'{\i}sica de la Universitat de Val\`encia \\
$^3$ Department of Physics and Astronomy, University of California, Irvine, CA 92697-4575, USA\\
$^4$ Las Cumbres Observatory, Goleta, California\\
}    

\begin{abstract} 

We review the present status of cosmological discoveries and how
these confirm our modern cosmological model, but at the same time we
try to focus on its weaknesses and inconsistencies with an
historical perspective, and foresee how the on-going big
cosmological projects may change in the future our view of the
universe.

\end{abstract}

\begin{quotation}
Motto: ``O dark dark dark. They all go into the dark,\\
The vacant interstellar space,
the vacant into the vacant" (T. S. Elliot, in ``East Coker", No. 2 of The Four Quartets)
\end{quotation}

\section{Introduction}

Cosmology is the study of the universe as a whole, it is the science
of the large-scale structure of the universe, its origin, and its
evolution from the early times into the future. In this context
universe means all that exists in a physical sense, not only the
part of the universe that we can observe \citep{ellis07} using our
telescopes and detectors on the ground and in space. The observable
universe could certainly be a tiny fraction of the whole universe,
even an infinitesimal fraction if the universe were infinite.

Cosmology today can be considered a branch of physics with a slight
difference: we cannot experiment with the subject of our discussion,
the universe, we can only observe it and model it.

The statement that ``we are living in the era of precision
cosmology" is certainly one of the most heard ones in the last 10
years in conferences, seminars and talks about the field. There is
no doubt that within this period, modern cosmology has expanded from
what Allan Sandage (1970) once described as ``the search for two numbers",
meaning the Hubble and the deceleration parameters. These and a
few more numbers are conforming now a self-consistent set, derived
from several different cosmological observations: high-redshift
supernovae, fluctuations of the Cosmic Microwave Background (CMB)
radiation, the large-scale structure of the universe, gravitational
lensing, etc., but the emergent concordance cosmology provided by
all these probes (sharing all beautiful hard-to-get cosmological
data) is in a sense disappointing. We need to claim for the
existence of gravitating non-baryonic dark matter of unknown nature,
and furthermore, the universe today has to be dominated by an exotic
dark energy, acting as a repulsive gravity. Some cosmologists take
those theories as seriously as Ptolemy and colleagues took epicycles
and deferents to reconcile the geocentric model with the early
observations of planetary motion, or how physicists prior to the
Michelson-Morley experiment considered the aether of undoubted
certain existence. Since cosmology is not an experimental science,
but an observational one, we must take this into account when we try
to falsify our theories in the sense advocated by Karl Popper (1959): our
requirement is just that our theories should be consistent with
present and future observations. In contrast to what happens in
experimental physics, astronomers cannot modify the object under
study. They cannot control it in any way; they only can observe it
many times, with different exposure times or at different
frequencies or observe many objects of the same type
\citep{kolb07a}. This fact inevitably conditions the way we do our
research and plan our observations.

\section{The establishment of the present cosmological paradigm}

Research done during the previous century established our Standard
Cosmological Model. Cosmology started to be considered a scientific
discipline with the introduction in 1917 of the General Theory of
the Relativity by Albert Einstein, which acts as theoretical
framework for the development of the cosmological models. In the
decade of 1920 Alexander Friedmann and Georges Lema\^{\i}tre suggested
solutions to the equations of Einstein that provided dynamic
universes, in expansion or in contraction. The discovery of the
expansion of the universe carried out by Edwin Hubble in 1929
allowed for non-static models of universe that accounted for the
observed expansion (the models of Friedmann-Lema\^{\i}tre that make use
of the Robertson-Walker metric).

The idea that the universe might experience constant change locally
and yet be on average invariable  for very long times or eternity
can probably be found among the ancient Greeks (Democritus). But the
modern version dates firmly to 1948 and a pair of papers by
\citet{bondi48}  and by \citet{hoyle48}, suggesting that the
expansion of the universe, as implied by Hubble's and later work
\citep{hubble29}, is perfectly real but that additional matter is
created at just the rate needed (about one atom per $10^6$ cm$^3$
per Hubble time) to keep the mean density constant, new galaxies
being constantly formed from that new matter. A few of the
observational objections to this picture would disappear if the new
material consisted of about 1 helium atom for every 10 hydrogens,
and perhaps 5 times as much in some form of dark matter capable of
gravitational (and perhaps weak) interactions only, though DM was
not, of course, part of the Bondi-Gold-Hoyle picture. Their
intentions were at least partially philosophical, for instance to
bring the process of creation within the observable universe.

The rate of creation required is very far below observability. But a
pure steady state universe requires that the average mass and
luminosities of galaxies, their clustering properties, and their
propensity to emit strong radio fluxes must not change with redshift
(time or place). The average age should be 1/3 of the Hubble time,
making our Milky Way unusually old (though critics who had not
thought through the issues tended to claim that the absence of young
galaxies was the greater objection). But   it was the requirement
for a constant percent of galaxies to be strong radio sources that
already cast serious doubt on the steady state model before 1960.
Counting radio sources  \citep{scheuer57,ryle61} was a disputed
issue, but after the discovery of the redshifts for quasars
\citep{schmidt63}, there was no doubt that strong radio sources had
been stronger and more common in the past \citep{schmidt68},
providing evidence against the steady state model.

It was also not easy to reconcile the apparent brightness and
angular diameters of distant galaxies with steady state requirements
(since average properties must not change with time), and if
creation was of pure hydrogen, then turning one-quarter of the
material to helium in stars implied galaxies 5-10 times brighter
than the ones we see.

In addition, clearly, there must not be anything found in the
universe that could only have arisen under conditions very different
from the present ones. Thus the 1965 discovery of the cosmic
microwave radiation \citep{cmb} was the death of the steady state model for most
astronomers who had not paid much attention to the earlier problems.
There remains a small group of supporters of a short
quasi-steady-state model, with much less of the simplicity enjoyed
by the original one and the need still to doubt the cosmological
nature of QSO redshifts as well as a need for intergalactic iron
filings to thermalize the CMB and colorlessly absorb light from
distant supernovae. We can admire their courage without having any
desire to follow their ideas.

The discovery of the cosmic background radiation emitted by the hot
gas when the universe was at 3000 degrees and had an age of about
380\,000 years was a definite support for a general acceptance of a
universe in expansion, with a finite age and an extremely dense and
hot beginning, in what the physicist Fred Hoyle pejoratively called
Big Bang. The name settled and the theory of the Hot Big Bang became
the basic cosmological model, with some of its predictions ending up
with clear observable successes, as the explanation (Alpher, Bethe \&
Gamow 1948; Peebles 1966) of the relative proportions observed in
our local environment of the light elements (helium, deuterium and
lithium).

The model itself was not exempt from some paradoxes, as the problem
of the extreme homogeneity and isotropy among parts of the universe
that had never been in causal contact (due to the finiteness of the
light speed) or did not provide a convincing reason to justify that
the density of matter and energy was so close to the critic value
(flatness problem). With the introduction of the concept of
inflation \citep{guth81} that suggests a phase of fast acceleration
of the cosmic expansion in the early stages of the universe, some of
these problems are solved, from the theoretical point, at the
expense of introducing an additional hypothesis, which certainly is
still not completely proved by observations.

\section{The dark side}
\subsection{Dark matter}

In the 1970s the need to advocate for the existence of a
considerable quantity of dark matter (DM) in the universe was clearly
established. The measurement of the Doppler shifts of star light in the external
parts of the spiral galaxies shows an unexpected behaviour: The
velocities of stars (or HII regions) orbiting around the galactic
center did not decrease following the foreseeable Keplerian
dynamical behaviour \citep{rubin70},  but instead remained roughly
constant to great distances from the galactic center. The presence
of dark matter in the galactic dynamics was used for rescuing the
works of Fritz Zwicky of the decade of 1930 from the oversight.
Zwicky had to advocate the existence of this type of matter ({\it
dunkle Materie}) to maintain the stability of the galaxy clusters
\citep{zwicky33}\footnote{Zwicky was not, however, the first either
to use the phrase dark matter or the first to report a number for
it.  James Jeans (1922) and Jacobus Kapteyn (1922)  estimate the
mass in the disk of the Milky Way (by method refined by Jan Hendrik
Oort in 1932), reporting the presence of dark stars.}. The
measurements of the average peculiar velocity dispersion in the
radial direction with values of the order of 1000 km/sec in the Coma
cluster led Zwicky to this conclusion. The velocities of galaxies
within the cluster are a consequence of the gravitational potential
associated to the total cluster mass. In this kind of virialized
systems, the potential energy is related with the kinetic energy,
--associated to the distribution of individual galaxy velocities--
through the virial theorem ($2K+U=0$), providing a method to
estimate the total cluster mass.

Other observations carried out in the 1980s, as the emission in
X-rays produced by the hot gas in clusters of galaxies or the image
distortions and magnifications produced by galaxy clusters acting as
gravitational lenses, have corroborated the need for dark matter.

It is essential to distinguish two aspects: existence and nature,
with the former quite firmly established and the latter much
constrained but still unknown. One can, in a sense, regard ``dark
matter" as a shorthand for a very large number of observations
on many scales, indicating that mass to
light ratios increase as you look at larger entities.  This was
pointed out in a pair of important and influential papers by
Einasto, Kaasik, and Saar (1974) and Ostriker, Peebles, and Yahil
(1974). These and other observations, when collectively plotted on a logarithmic scale of
luminosity-to-mass ratio vs. the length scale, show a
monotonic rise from unity for 1-parsec diameter young star clusters
to something like 200-300 $M_\odot/L_\odot$ for the largest superclusters of
galaxies and other very large scale structures explored with weak
gravitational lensing. The rise does not continue on larger scales,
though many back in the 1980s thought it would. Such a plot could
have been made before the Second World War, using Hubble's numbers
for the inner parts of galaxies, Babcock's rotation curve for M31
\citep{babcock39}, Holmberg's binary galaxies \citep{holmberg40},
and the data on the Coma and Virgo clusters from \citet{zwicky33}
and \citet{smith36}.

More modern data include a still large range of systems -- disks of
galaxies from motions of stars and gas perpendicular to them, whole
clusters from X-ray and lensing data, and the very largest scale
information we have from the CMB, Type Ia supernovae, and weak
gravitational lensing. The only possible conclusions are either that
gravity becomes monotonically stronger on large scales or that the
ratio of non-luminous matter increases with length scale. The latter
is by far the majority view in the astronomical community and
centers around something like 23\% of the closure density being in
non-luminous, non-baryonic dark matter.

A number of ideas in modern physics imply dark matter candidates, of
which the most often  sought is supersymmetric partners of known
particles, the lowest-mass supersymmetric particle in 4-d space time
or perhaps the lowest-mass Kaluza-Klein particle in 5-d space time.
Current observations and experiments are looking for three
manifestations: (1) photons or $e^{\pm}$ pairs produced when DM
particles annihilate today, (2) scattering of the particles in large
laboratory detectors (made of NaI crystals, very pure water, or
other substances), and (3) production of DM particles in
accelerators like the upcoming LHC. Other viable DM candidates
include axions, black holes in a few unprobed mass ranges,
topological singularities, and many more exotic entities.
Remembering here, as in other places in this chapter, that theories
are cheap but telescopes or accelerators are expensive, we encourage
our theoretical colleagues to think broadly and to deduce possible
detectable consequences of their DM-candidates, particularly
consequences that might be found (like gamma ray emissions or
positron excesses) in projects that are being carried out for other
purposes. Very large investments in programs narrowly aimed at a
single candidate are harder to feel positive about \citep{white07} .

\subsection{Dark energy}

Dark energy (DE), like dark matter, is a shorthand for a large number of
observations and ideas.  But in this case, an idea came first.  The
differential equations for a homogeneous, isotropic, relativistic
universe are second-order, and so admit two integration constants.
The first (in suitable units) is the Hubble parameter at some
reference time. The second takes the form of a uniform density
(always positive) and pressure (which can be positive or negative),
with negative pressure tending to oppose ordinary gravity \citep{mcvittie56}.  Einstein
called it $\lambda$ and wanted it initially to permit a static
universe (which turned out to be unstable).  It is now generally
written as $\Lambda$ , and Einstein left it out of his publications
after 1930.  In 1934, however, R.C. Tolman included the possibility
of both positive and negative values of $\Lambda$, and one of his
model universes, with negative pressure $\Lambda$, expanded from a
singularity to infinite size, with an empty de-Sitter universe as
its limit.

Despite the frequent phrases ``Einstein's infamous cosmological
constant" and ``Einstein's worst blunder," $\Lambda$ has never
entirely disappeared from the literature, serving in at least a few
minds as a solution to the problem presented by a universe somewhat
younger than its contents, a problem never entirely eliminated by
recalibrations of the Hubble constant between 1952 and the present.
De Vaucouleurs, for instance, always included $\Lambda$ in his
cosmological discussions, beginning in about 1956.  There was
another revival around 1970 in connection with the apparent excess
of QSOs with redshifts close to 1.95.  Eventually regarded as a
selection effect, this could, in principle, have been a signature of
a coasting phase in an open universe with non-zero $\Lambda$.
Incidentally, the critical density case (now thought to be very
close to reality) has no coasting phase, only an inflection point in
the expansion parameter $a(t)$.

Observational cosmology, gradually involving many more kinds of
observations than just Sandage's ``search for two parameters"
proceeded apace, and by the time of the 1997 IAU General Assembly in
Kyoto, evidence had accumulated from large scale structure, galaxy
formation simulations, ages, and big bang nucleosynthesis for a flat
(critical density) universe with something like $70\%$ of the
gravitation coming from negative-pressure $\Lambda$.  Since then,
the numbers favoured by several panel members there (4-5\% baryons,
$23-25\%$ dark matter, and the rest $\Lambda$) have been reinforced
by results of studies of weak gravitational lensing, supernovae, and
angular fluctuations of the CMB seen by WMAP.

For many decades cosmologists have been trying to  quantify how the
expansion of the universe discovered by \citet{hubble29} was slowing
down due to gravity.  However, in 1998, two independent teams
\citep{riess98,perlm99} presented convincing evidence for just the
contrary: an accelerated expansion. They used high-redshift Type Ia
supernovae (SNe Ia) as standard candles \citep{1993ApJ...413L.105P}.
The behaviour of its calibrated luminosity-distance as a function of
the redshift  of their host galaxies ruled out the Einstein-de
Sitter spatially flat cosmological model, indicating that the cosmic
expansion had been speeding up during the last 5 Gyr or so.
$\Lambda$ was then definitively rescued from the wastebasket in
1998 with the interpretation of the luminosity-distance-redshit
relation of very distant type Ia supernovae as evidence for
acceleration in cosmic expansion. The two mentioned teams analyzed  a
set of high-$z$ supernovae and found them fainter than expected.
After ruling out possible systemic obscuration by dust or
evolutionary effects, they interpreted the dimmer luminosity as a
consequence of being farther away, and thus implying an acceleration
in the expansion.

At this point, physicists step into the picture, asking ``what is
$\Lambda$ apart from the integration constant that Einstein called
it?\footnote{In a letter to Besso quoted by \citet{kragh96},
Einstein explained: ``Since the universe is unique, there is no
essential difference between considering $\Lambda$ as a constant
which is peculiar to a law of nature or as a constant of
integration."}"  And ``why does it have the numerical value we
find?"  New words, especially dark energy and quintessence, are
invented to describe it and to suggest the possibility of variation
with time and perhaps space.  It acquires an equation of state:
$p=w\rho$, where $w$ exactly and always $-1$ is just $\Lambda$
back again, therefore the simplest form of dark energy is the
stress-energy of empty space --the vacuum energy--, which is
mathematically equivalent to the Einstein's cosmological constant,
but other values of $w$ and time variability might allow eventually
recontraction of the universe or expansion so fast that it tears.
These other forms of dark energy that dynamically evolve with time
have been considered in the literature  \citep{peebles03} and are called
``quintessence". The astronomical community has embraced very
quickly the idea of accelerated expansion. The solid arguments
accompanying the observations of the SNe Ia have been confirmed with
spectroscopic analyses \citep{bronder08,sullivan09} that test for
possible systematic uncertainties. Their results confirm the
reliable use of SNe Ia as standardized candles. Moreover, there
exists other independent observational evidence supporting  the
accelerated expansion of the universe. For a review see
\citet{2008ARA&A..46..385F}. Amongst these probes, one of the most
promising techniques is the measurement of the baryon acoustic
oscillations (BAOs) in the large-scale distribution of matter in the
universe \citep{eisens05,cole05,martinez08}.

Dark energy in this modern sense has been associated with the last
gasp of inflation, new scalar fields, vacuum field energy, and other
innovative physics that we do not pretend to fully understand.  The
catch in most cases is that the natural amount should have a density
of one Planck mass ($10^{-5}$ g) per Planck volume ($10^{-99}$
cm$^3$), something like $10^{120}$ larger than the $73\%$ of closure
density implied by the concordant observations of supernovae, the
CMB, large scale structure, etc.

\section{Falsifiability, confirmability, evidence and all that}

Belief in a scientific theory must always be established on an objective
assessment of the evidence. That several lines of evidence give the
same numbers is not a perfect guarantee of correctness -- Kelvin was
sure he knew the age of the sun and solar system because his
calculation of the cooling age of the earth gave the same 10-20 Myr
as the lifetime of a solar mass star with gravitational contraction
as its only energy source.

There are respectable motivations to take the $\Lambda$-CDM model seriously as
a hypothesis about the universe, but this is not equivalent to
declaring its unvarnished truth.

\subsection{Falsifiability}

For many years, the Einstein-de Sitter model was  the most popular
hypothesis for a dynamical description of the universe. The high redshift
Type Ia supernovae were a strong evidence supporting its
inconsistency. Today the evidence against this once favoured
hypothesis comes from many different observations.

But the important thing of the present standard model ( $\Lambda$-CDM) is that it can
make predictions that can be tested by observations and therefore
the theory is falsifiable (vulnerable to being shown false by observation or
experiment).  An example from the past in cosmology: steady
state was falsified (fairly quickly, in fact, as we had already
explained) because it made some definite predictions. An example for
hopefully the near future: of the popular ideas out there now,
inflation is surely falsifiable

a) via polarization structure of CMB and such, indeed it is looking
a little weak in the knees now: polarization-sensitive CMB
experiments will come very soon.

b) via detection of a stochastic gravitational wave background.

\subsection{Consensus?}
The consensus about the existence of dark matter is high. The
evidence of its existence is clearly stronger than the evidence of
the existence of dark energy. Prospects for the detection of dark
matter candidates are ongoing in different experiments.  There are
interesting, well-motivated DM candidates (and also of course some
silly ones), being the neutralino the everyone's favourite candidate
for the moment \citep{kolb07b}.

One of the first alternatives to dark matter was formulated by \citet{finzi63}
to guarantee the stability of clusters of galaxies without advocating for dark matter. 
Finzi's  hypothesis was a modification of the gravitational Newton 
law in such a way that the actual attraction at long distances should be stronger than the
value predicted by the Newton's Law, but probably the optimal version
of this has not yet been put forward. Two decades later, Milgrom (1983) proposed a different
alternative to the dark matter based in a modification of Newtonian dynamics (or MOND for short). 
In this hypothesis, the Newton's second law of dynamics is modified in such a way that
when accelerations experienced by objects are smaller than a certain value, the gravity 
force is inversely proportional to the distance, instead of to the distance squared. This modification 
explains rather well the flat rotation curves of the spiral galaxies \citep{sanders02} 
whose dynamics are a consequence of the luminous baryonic matter alone, with no 
need to claim for dark matter.
Although MOND has successfully explained other cosmological observations, it does not 
reproduce so well the dynamics of clusters of galaxies and the observations of weak and strong lensing and the CMB.

\subsection{Success?}

Should we regard the ``discovery" of the dark energy and the
acceleration of the expansion in the universe as a scientific
success? Certainly in 1998, Science magazine considered this
discovery as the breakthrough of the year and we agree with that
decision. This discovery put together many astrophysicists,
cosmologists and high-energy physicists in a common effort trying to
understand the nature of the dark energy (see for example the  Dark
Energy Task Force report by \citet{detf} and the ESA-ESO Working Group 
on ``Fundamental Cosmology by \citet{peacock06}). But is the discovery of
the dark energy by itself a scientific success? It is certainly a
crucial step, but probably the story should not be considered a
success at least until it can be well explained in terms of an
existing theory. As Lee Smolin (2006) says: ``The discovery of the dark
energy cannot be counted as success, for it suggests that there is a
major fact that we are all missing." Of course,
this statement does not subtract the merit to the Supernova
Cosmology Project led by Saul Perlmutter and the High-z Supernova
Search Team led by Brian Schmidt and other observations supporting
the accelerating world models; what it means is that the presence of
a non-zero vacuum energy is a problem that has to be explained in
much the same way the existence of the aether was a problem  that
had to be explained by the physicists prior to the Michelson-Morley
experiment. In that case the experiment acted denying the existence
of the aether and that was the solution. In Cosmology, future
planned and ongoing observations have as a major scope to understand
the nature of the dark energy \citep{jdem}. Some of these projects
are based on distant supernovae \citep{wood07} and BAOs
\citep{benitez09,wigglez,boss,space}.

In any case and if what you care about is things like galaxy
formation and evolution, then DE was not important when most of the
relevant processes were going on. As \citet{white07} has remarked in
a recent essay DE is an interesting problem to plan astronomical 
observations, but it is just ``one of many."

\section{Conclusions}

It seems clear that  many of the pre-Copernican astronomers who made
earth-centered models gradually more complex to match better
observations thought -- according to historians, anyhow -- that they
were describing the phenomena, not explaining them. Are cosmologists
continuously re-editing an undeclared unsuccessful model of the
universe to accommodate it to new and unexpected observations?
\citep{disney07}.   Several authors \citep{horvath} are already
declaring the crisis of the present cosmological model and
advocating for the need of a paradigm shift in a Khunian sense
\citep{khun62} but, at the same time, the general adherence to the
mainstream concordance $\Lambda$-CDM model does not leave too much
room for thinkers outside the accepted cosmic paradigm.

Does this mean that theorists or observers or both should give up on
the universe and go back to studying cataclysmic variables (of which
we are secretly very fond)?  Certainly not!  What it does mean, we
think, is

1. Observers should be careful when combining many different sorts
of data into a many-parameter model that they have not started off
their minimization process from a place in the associated
many-dimensional space that will trap them in a false minimum of
values that seem to be the best possible fit but are far from the
truth.

2. Theorists should put forward as many candidates as they want, but
should ask whether their favorites (for instance $w$ a smidge larger
or smaller than $-1$) might have observable/testable consequences that
can be extracted from programs and missions that have significant
potential for learning other important things about the universe and
its contents if the dark energy continues, as it has done so far, to
act precisely like a pure, infamous cosmological constant.  This, of
course, especially true for candidates associated with various
multiverse concepts.

As Rocky Kolb (private communication) has emphasised after reading a first draft
of this manuscript:  ``Our goal must not be a cosmological model that just explains the
observations, the ingredients of the cosmological model must be
deeply rooted in fundamental physics.  Dark matter, dark energy, modified
gravity, mysterious new forces and particles, etc., unless part of
an overarching model of nature, should not be part of a cosmological
model.  We may propose new ideas, but they
must wither unless nourished by fundamental physics."

\acknowledgements 

We thank Rocky Kolb, Jos\'e Adolfo de Azc\'arraga, Ramon Lapiedra,
Mar\'{\i}a Jes\'us Pons-Border\'{\i}a, and Alberto Fern\'andez-Soto
for many comments and suggestions. This work has been supported by
the Spanish Ministerio de Ciencia e Innovaci\'on projects ALHAMBRA
(AYA2006-14056) and PAU (CSD2007-00060), including FEDER
contributions.

\end{document}